\documentclass{jaa}

\usepackage{graphicx}
\usepackage{graphicx}
\usepackage{xcolor}
\newcommand\farcdeg{\mbox{$.\!\!^\circ$}}%
\newcommand\farcsec{\mbox{$.\!\!^{\prime\prime}$}}%
\newcommand\etal{{\em et al. }}%

\begin{document}\sloppy


\title{ Optical Observations of star clusters NGC 1513 and NGC 4147; white dwarf WD1145+017 and $K$ band imaging of star forming region Sh2-61 with the 3.6 meter Devasthal Optical Telescope}


\author{Ram Sagar\textsuperscript{1,2}, R.K.S. Yadav\textsuperscript{2,*}, S.B. Pandey\textsuperscript{2},
Saurabh Sharma\textsuperscript{2}, Sneh Lata\textsuperscript{2} and Santosh Joshi\textsuperscript{2}}
\affilOne{\textsuperscript{1}Indian Institute of Astrophysics, Sarajapur road, Koramangala Bengaluru, 560 034, India.\\}
\affilTwo{\textsuperscript{2} Aryabhatta Research Institute of Observational Sciences, Manora Peak Nainital, 263 001, India.}


\twocolumn[{
\maketitle
\corres{rkant@aries.res.in}
\msinfo{2021}{XX}

\begin{abstract}
 The $UBVRI$ CCD photometric data of open star cluster NGC 1513 are obtained with the 3.6-m Indo-Belgian Devasthal optical telescope (DOT). Analyses of the GAIA EDR3 astrometric data have identified 106 possible cluster members. The mean proper motion of the cluster is estimated as $\mu_{\alpha}Cos{\delta}=1.29\pm0.02$ and $\mu_{\delta}=-3.74\pm0.02$ mas yr$^{-1}$.  Estimated values of reddening $E(B-V)$ and distance to the NGC 1513 are 0.65$\pm$0.03 mag and 1.33$\pm$0.1 kpc respectively. An age of $225\pm25$ Myr is assigned to the cluster by comparing theoretical isochrones with deep observed cluster sequence. Using observations taken with the 3.6-m DOT, values of distance and age of the galactic globular cluster NGC 4147 are estimated as $18.2\pm0.2$ Kpc and $14\pm2$ Gyr respectively. The optical observations of planetary transit around white dwarf WD1145+017 and $K$-band imaging of star-forming region Sharpless Sh 2-61 demonstrate observing capability of 3.6-m DOT. 
 
 Optical and near-infrared observations of celestial objects and events are being carried out routinely with the 3.6-m DOT. They indicate that the performance of the telescope is at par with those of other similar telescopes located elsewhere in the world. We therefore state that this observing facility augurs well for multi-wavelength astronomy including study of astrophysical jets. 
\end{abstract}
}]
\doinum{12.3456/s78910-011-012-3}
\artcitid{\#\#\#\#}
\volnum{000}
\year{0000}
\pgrange{1--}
\setcounter{page}{1}
\lp{1}

\section{Introduction}

The 3.6-m Indo-Belgian Devasthal optical telescope (DOT) is located at the mountain peak (longitude = 79\farcdeg{7} E, latitude = 29\farcdeg{4} N, altitude = 2424$\pm$4 m) of Devasthal (meaning 'abode of God') in Nainital district of Kumaun region, Uttarakhand. The location was identified after decades of detailed site survey using modern instruments (Sagar \etal 2000; Stalin \etal 2001 and references therein). The telescope and its performance are described in details by Kumar \etal (2018); Omar \etal (2017) and Sagar \etal (2019a, b, 2020). 


 Being a joint venture between India and Belgium, the 3.6-m DOT was technically activated jointly by premiers of both countries from Brussels, Belgium on March 30, 2016. Since then, the telescope is used  routinely for carrying out optical and near-infrared (NIR) observations of galactic and extra-galactic celestial objects as well as transient astronomical events (Pandey \etal 2016, 2021; Kumar \etal 2020, 2021a,b, 2022; Aryan et al. 2021; Gupta et al. 2021, 2022a,b). Presently, there are 4 back-end instruments at the 3.6-m DOT. Two of them namely 4K$\times$4K CCD Imager (Pandey \etal 2018; Kumar \etal 2022) and ARIES Devasthal faint object spectorgraph Camera (ADFOSC) (Omar \etal 2019a, b, c) are used for optical observations while TIFR Near Infrared Imaging Camera-II (TIRCAM2) and TIFR-ARIES Near Infrared Spectrometer (TANSPEC) are used for NIR observations (Ojha \etal 2018; Baug \etal 2018). Both optical and NIR observations taken with these instruments have shown that angular resolutions and detection limits achieved with the telescope are at par with similar telescopes located elsewhere in the globe (Sagar \etal 2019a, b, 2020 and references therein). 

   In this paper, we present CCD Optical broad band photometric and $K$ band observations taken with the 3.6-m DOT. Star cluster NGC 1513 and NGC 4147 and white dwarf WD145+07 were observed using CCD Imager while TIRCAM2 was used for $K$ band imaging of star forming region SH 2-61. These observations and multi-wavelength archival data are used to derive precise astrophysical parameters of the star clusters under study. The outline of the paper is as follows. Sections 2 to 5 are devoted on the open star cluster NGC 1513. The optical observations of cluster NGC 4147; planetary transits around WD1145+017 and deep $K$-band imaging of a star forming region Sharpless Sh 2-61 are discussed in sections 6, 7 and 8 respectively. The summary and conclusions are presented in the last section.

\section{The open star cluster NGC 1513}

The member stars in an open star clusters are loosely concentrated and gravitationally bound to each other. The GAIA EDR3 kinematical data are very important to interrogate the motion of open star clusters. The proper motion and parallax data are being frequently used to identify the cluster members as they play very crucial role in deriving the basic parameters of the clusters (Yadav \etal 2008, 2013).

   The open cluster NGC 1513 is located in Perseus arm of the Milky way. Its equatorial and galactic coordinates are $\alpha_{2000} = 04^{h}09^{m}98^{s}$, $\delta_{2000}=49^{\circ} 31^{\prime} 00^{\prime\prime}$ and $l$=152$^\circ$.6, $b$=-1$^\circ$.57 respectively. This cluster is classified as II2m, a moderately rich cluster with little central concentration by Trumpler (1930). Bronnikova (1958) determined the proper motion and membership probability using photographic plates having an epoch difference of $\sim$ 55 years. Radius of the cluster is estimated to be $\sim 7^{\prime}$ (see Table \ref{lItrature}). Barkhatova \& Dryakhlushina (1960) and del Rio \& Huestamendia (1988) investigated the cluster by carrying out photographic and photoelectic photometry of 49 and 116 stars respectively. Frolov \etal (2002) studied this cluster astrometrically and identified 33 probable members of the cluster. Using CCD $BV$ data, cluster parameters have been determined by Maciejewski \& Niedzielski (2007). All these information are tabulated in Table \ref{lItrature}. They indicate that cluster age is $\sim$ 100 Myr while distance is $\sim$ 1.3 Kpc. 
   

\begin{table*}
\centering
\caption{Existing relevant studies of NGC 1513 are listed in chronological order. In the first column, studies based on Photographic, Proper motion, Photoelectric and Charge coupled devices are abbreviated as PG, PM, PE and CCD respectively. Letter t in column 6 denotes age (in years). Limiting magnitude of the study is denoted by Lm in column 7. In the last column, Number 1, 2, 3, 4 and 5 denote Bronnikova (1958), Barkhatova \& Dryakhlushina  (1960), del Rio \& Huestamendia (1988), Frolov \etal (2002) and  Maciejewski \& Niedzielski (2007) respectively.}
\begin{tabular}{llc ccc lc} \hline \footnotesize
 Study & pass bands & Radius & Distance & $E(B-V)$ & log(t) & Lm & Reference   \\
          &       & (arc-minute) & (parsec) & (mag) &  & (mag) &  \\ \hline
PG, PM    & $m_{pg}$  &    &   &    &  & $m_{pg} \sim$ 14 & 1 \\
 PG photometry & $m_{pg}$, $m_V$ & 7 &860 & &  & $m_{pg} \sim$ 16.5 & 2 \\
PE, PG photometry & $UBV, RGU$ & 7 & 1320 & 0.53 & 8.18 & $G \sim$ 16.5 & 3 \\  
PM, CCD photometry & $BV$ &   & 1320 & 0.67 & 8.4 & $V \sim$ 16.5 & 4 \\
 CCD photometry &$BV$ & 9.2 & 1320 & 0.76 & 7.4 & $V \sim$ 19 & 5 \\ \hline
 \end{tabular}
\label{lItrature}
\end{table*}

As evident from the above literature surveys and Table \ref{lItrature}, NGC 1513 lacks  multi-wavelength deep $UBVRI$ CCD photometric study. Also, existing studies are mainly limited to $V \sim$ 16 mag and hampered by field star contamination. A detailed photometric and kinematical studies of this poorly studied intermediate age star cluster NGC 1513 are carried out below by combining present $UBVRI$ CCD photometric data with the GAIA EDR3 high precision proper motion data. 

\begin{figure}
\begin{center}
\includegraphics[width=8cm, height=8cm]{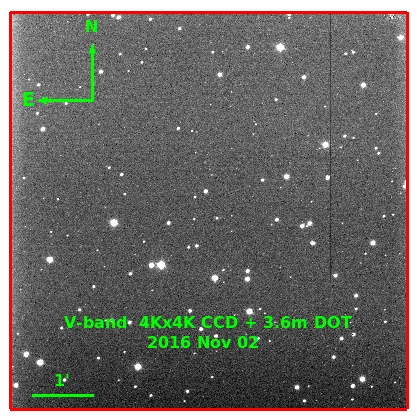}
\caption{ $V$ image of the cluster NGC 1513 taken with 3.6-m DOT. North is up and east is left. }
\label{id_open}
\end{center}
\vspace{-1.0cm}
\end{figure}

\subsection{Observations and data reductions}

NGC 1513 was observed in $UBVRI$ photometric passband on 2$^{nd}$ and 3$^{rd}$ November 2016 with the 3.6-m DOT. Images were collected with $4K\times4K$ CCD camera in 4$\times$4 binning mode. At the f/9 Casssegrain focus of the telescope, a 15 $\mu$m pixel size of the CCD camera covers 0$^{\prime\prime}$.1 on the sky and field-of-view of the camera becomes 6$^{\prime}.5\times6^{\prime}.5$. The gain and readout noise of CCD is 5 e$^{-1}/ADU$ and 10 e$^{-1}$ (Pandey \etal 2018; Kumar \etal 2022) respectively. Log of observations is listed in Table \ref{log}. Figure 1 shows the observed cluster image in $V$ filter. We adopted the standard procedure for raw-data processing in the IRAF\footnote{IRAF is distributed by the National Optical Astronomical Observatory which are operated by the Association of Universities for Research in Astronomy, under contract with the National Science Foundation} environment. The DAOPHOT/ALLSTAR (Stetson 1987, 1992) software was used to do the photometry. The photometry was done on individual images. The instrumental magnitudes were derived using quadratically varying Point Spread Function (PSF). The magnitudes was aligned to that of a deepest frame for each filter and the final catalogue was created including all the objects identified at least in two filters.

The Landolt standard field PG0231 (Landolt 1992) was observed to transfer the instrumental magnitudes into standard values. For this, we used six standard stars.  Their range in brightness and colours are 12.77 $\le V \le$ 16.10 and $-0.34 < (B-V) < 1.44$ respectively. The transformation equations obtained are as follows: \\

\noindent $u$=$U$+3.50$\pm0.01-(0.09\pm$0.01)($U-B$)+(0.49$\pm$0.02)X

\noindent $b$=$B$+1.66$\pm0.01-(0.09\pm$0.01)($B-V$)+(0.32$\pm$0.01)X

\noindent $v$=$V$+1.58$\pm0.01+(0.04\pm$0.01)($B-V$)+(0.21$\pm$0.01)X

\noindent $r$=$R$+1.51$\pm0.01+(0.09\pm$0.01)($V-R$)+(0.13$\pm$0.01)X

\noindent $i$=$I$+1.72$\pm0.01-(0.04\pm$0.02)($V-I$)+(0.08$\pm$0.01)X \\

where $u, b, v, r$  and $i$ are the aperture corrected instrumental magnitudes and $U, B, V, R$ and $I$ are standard magnitudes whereas $X$ is the air mass. For the atmospheric extinction coefficients, we assumed the typical values for Devasthal site (Kumar \etal 2022). The errors in zero points and colour coefficients are $\sim$ 0.01 mag. The $X$ and $Y$ coordinates of the stars in the observed region are converted to Right Ascension (RA) and Declination (Dec) of J2000. To get the astrometric solution, we used the SkyCat tool and Guide Star catalogue v2 (GSC-2). We considered 100 bright stars for which we have both RA and DEC and the corresponding pixel coordinates. By using the task $CCMAP$ and $CCTRAN$ under IRAF, we converted the pixel coordinates into RA and DEC of all the stars  observed in the field. The $U, B, V, R$ and $I$ magnitudes of all stars observed in the region of NGC 1513 are listed in Table 3 along with RA, DEC and proper motion membership probability.A sample of this table is given here while full table is available online  as well as from the corresponding author.    

\begin{figure}
\vspace{-1.6cm}
\begin{center}
\includegraphics[width=8cm, height=8cm]{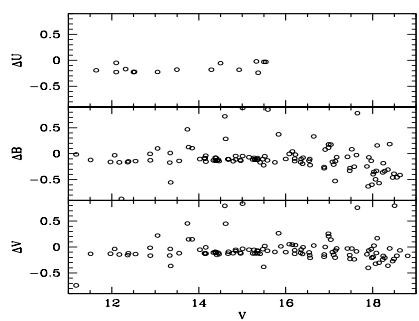}
\vspace{-2.06cm}
\caption{ Comparison of photometric $U, B$ and $V$ magnitudes. $B$ and $V$ CCD magnitudes are taken from Maciejewski \& Niedzielski (2007) while photoelectic $U$ magnitudes are taken from de Rio \& Huestamendia  (1988). The difference in magnitudes ($\Delta$U, $\Delta$B and $\Delta$V) are the present minus literature values.}
\label{comp}
\end{center}
\vspace{-1.0cm}
\end{figure}

\begin{figure*}
  \includegraphics[width=9cm, height=9cm]{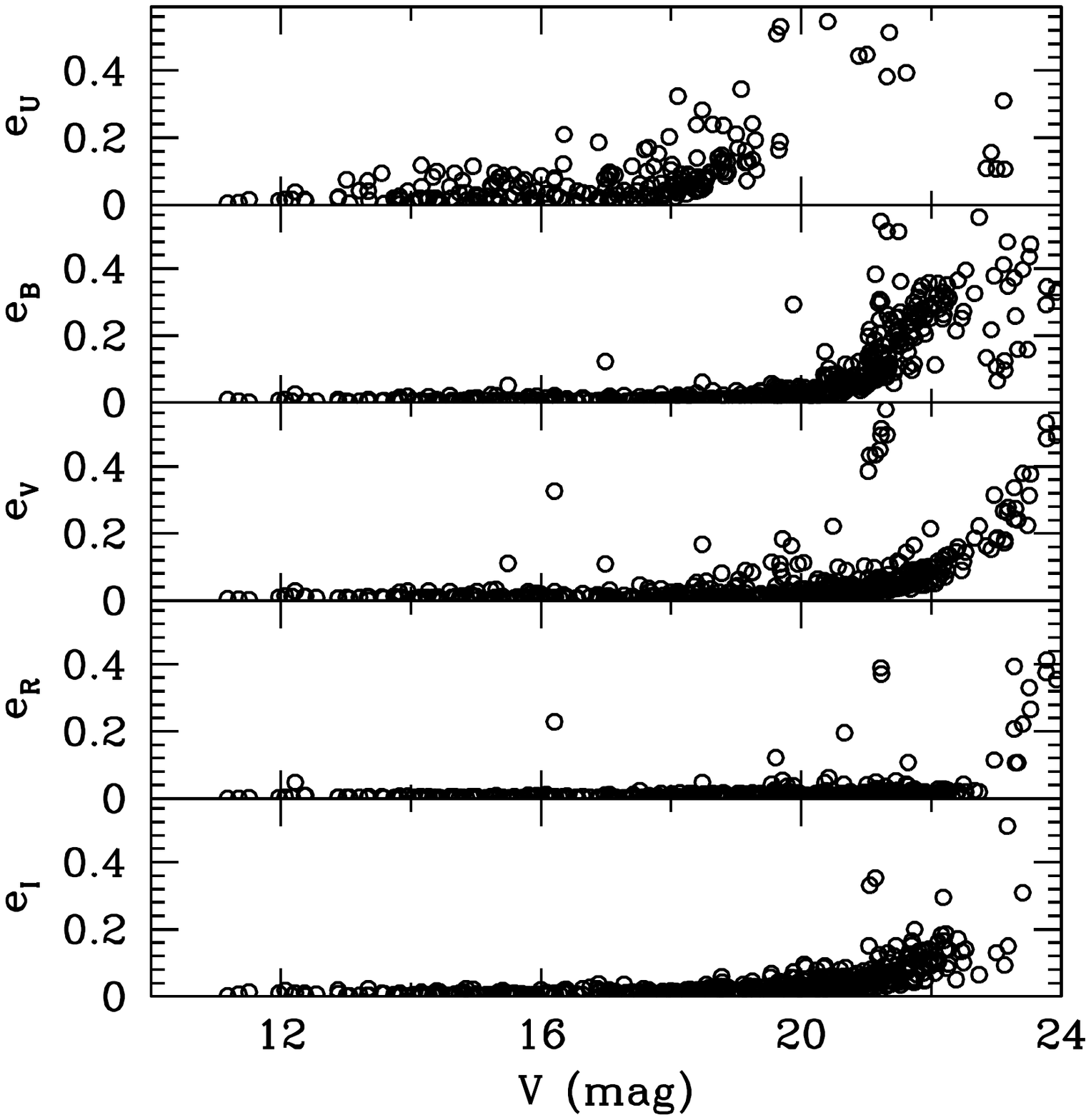}
 \includegraphics[width=9cm, height=9cm]{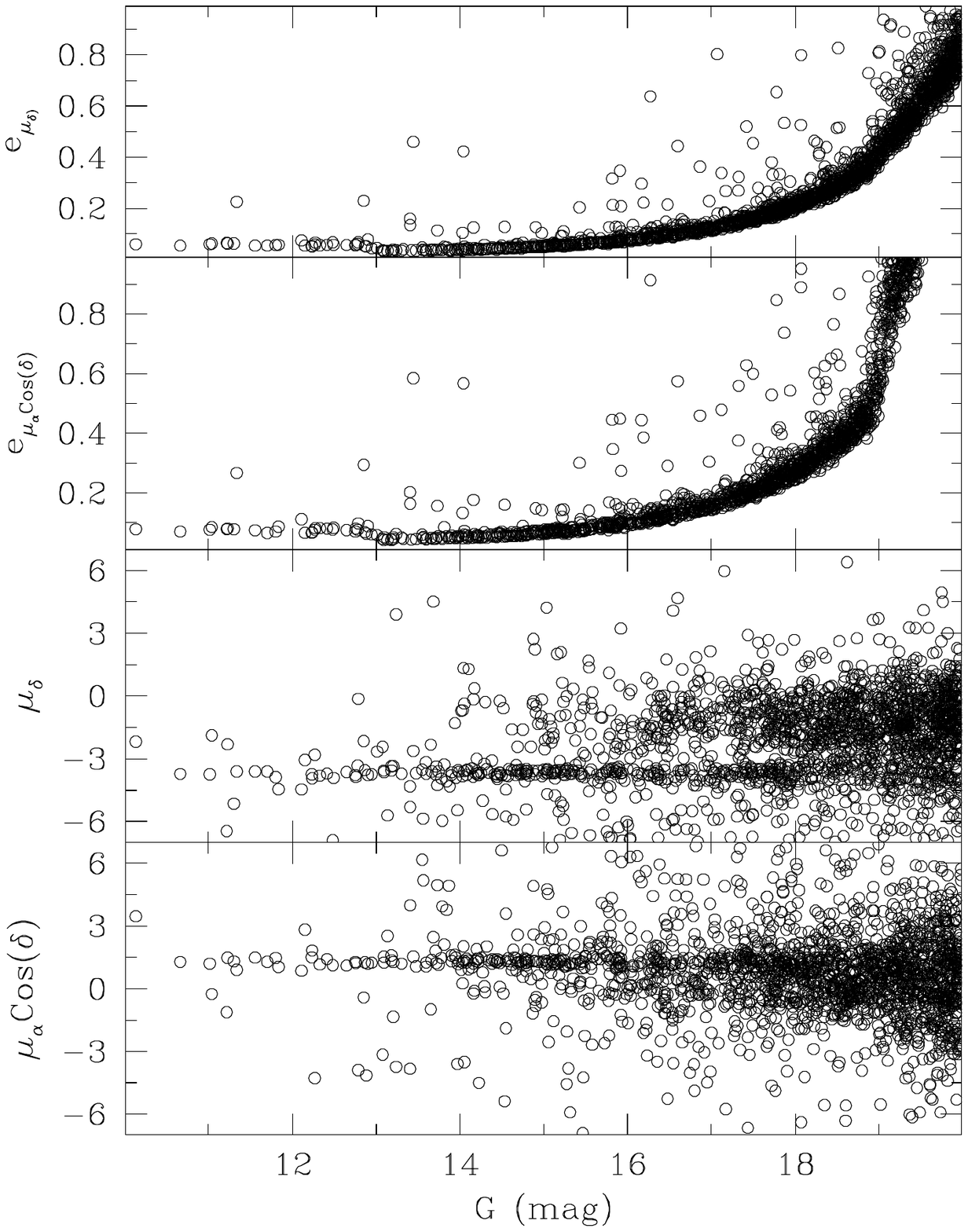}
\vspace {-2.6cm}
\caption{(Left panel) DAOPHOT errors in $U,B,V,R$ and $I$ band against $V$ magnitude.  (Right panel) GAIA EDR 3 catalogue proper motion $\mu_{\alpha}Cos{\delta}$ and $\mu_{\delta}$ with their errors against $G$ mag.}
\label{error}
\end{figure*}

Fig. \ref{comp} shows comparison of the present photometry with $B$ and $V$ CCD data from  Maciejewski \& Niedzielski (2007) and photoelectric $U$ data from de Rio \& Huestamendia (1988). We have cross identified the photometric catalogues by assuming that stars are correctly matched if the difference in position is less than 1 arc-sec. In this way, we identified 103 common stars in $B$ and $V$ and only 13 stars in $U$. The mean values of difference between two photometries are $-0.07\pm0.06$, $-0.07\pm0.08$ and $-0.14\pm0.08$ mag in $V$, $B$, and $U$ filters respectively. This indicates that present magnitudes are in fair agreement with literature except for $U$ mag where the difference is large.

The left panel of Fig. \ref{error} shows the internal errors in $U$, $B$, $V$, $R$ and $I$ magnitudes derived from DAOPHOT against $V$ magnitude. The mean errors are $\le$ 0.03 mag at $V$ $\sim$ 20$^{th}$ mag for $B$, $V$, $R$ and $I$ magnitudes while they are $\le$ 0.04 mag for $U$ magnitude at $V$ $\sim$ 18$^{th}$ mag. 



\begin{table}
\centering
\caption{ Log of optical observations of NGC 1513 are given along with exposure times in each filter.}
\begin{tabular}{clllll}
\hline
Date  &\multicolumn{5}{l}{Exposure Time (in seconds) in filter}   \\ \cline{2-6}  
      & $U$ & $B$ & $V$ & $R$ & $I$ \\ \hline
2 Nov 2016 & 300$\times$2 & 300$\times$6 & 300$\times$2 & 300$\times$6 & 10$\times$2 \\ 
3 Nov 2016 & 30$\times$2 & 30$\times$2 & 10$\times$2 & 10$\times$3 \\
\hline
\end{tabular}
\label{log}
\end{table}

\section{Proper motion data of NGC 1513}
The kinematical data of clusters are very useful for separating cluster members from field stars and also to estimate the mean proper motion of the cluster. We, therefore, downloaded the kinematical data from GAIA EDR3 catalogue (Gaia Collaboration \etal 2018) within the radius of $\sim$ 10$^{\prime}$ around the center of NGC 1513. It provides positions on the sky $(\alpha, \delta)$, parallaxes and proper motions ($\mu_{\alpha} cos\delta, \mu_{\delta}$) with limiting magnitude of $G=21$ mag. The error in parallax is $\sim$ 0.04 milli-arc-second (mas) for stars brighter than $G\sim15$ mag while for stars of $G\sim17$ mag, it is $\sim$ 0.1 mas. The proper motions and their corresponding errors for the cluster NGC 1513 are plotted against $G$ magnitude in the right panel of Fig \ref{error}. The errors in the corresponding proper motion components are $\sim$ 0.05 mas/yr for $G\le15$ mag), $\sim$0.2 mas/yr for $G\sim17$ mag and $\sim$1.2 mas/yr for $G\sim20$ mag.

\subsection{Mean proper motion}

\begin{figure}
\vspace{-3.8cm}
\begin{center}
\includegraphics[width=10cm, height=12cm]{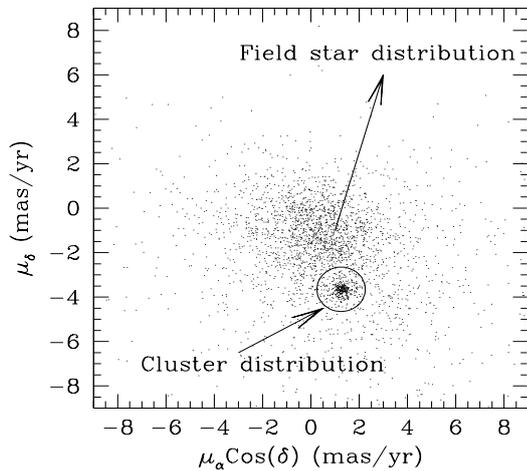}
\vspace{-3.6cm}
\caption{ Proper motion vector-point diagram (VPD) of the stars in the region NGC 1513. The distributions of cluster members and field stars are shown with arrows. Both distributions are clearly visible in the VPD.}
\label{vpd}
\end{center}
\vspace{-1.0cm}
\end{figure}

In Fig. \ref{vpd}, we show the vector-point diagram (VPD) of stars located in the region of NGC 1513 using the proper motion data in right accession ($\mu_{\alpha}$Cos$\delta$) and declination ($\mu_{\delta}$) directions. An inspection of the VPD exhibits two groups of stars. Dense population of stars shows the cluster members distribution while the scattered distribution represents the field stars. To select the probable cluster members, only stars located within a circle of 1 mas/yr radius shown in Fig. \ref{vpd} are considered. The radius of the circle is chosen as the best compromise between losing members with poor proper motions, and including field stars.

\begin{figure}
\vspace{-8.5cm}
\begin{center}
\includegraphics[width=10cm, height=12cm]{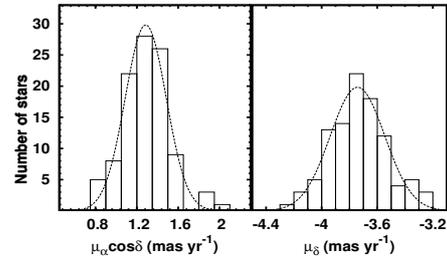}
\vspace{-1.80cm}
\caption{The proper motion histogram in $\mu_{\alpha}$Cos$\delta$ and $\mu_{\delta}$ in a bin of 0.1 mas/yr. The Gaussian function is fitted to the histograms to the central bins.}
\label{hist}
\end{center}
\vspace{-1.0cm}
\end{figure}

To estimate mean proper motion of the cluster, we plot histograms of the above selected cluster members in both $\mu_{\alpha}$Cos$\delta$ and $\mu_{\delta}$ directions in  Fig \ref{hist}. The fitting of Gaussian function to the histograms, provides the mean proper motions as $1.286\pm0.015$ and $-3.742\pm0.016$ mas/yr in $\mu_{\alpha} cos{\delta}$ and $\mu_{\delta}$ respectively. These precise estimates agree very well with the corresponding values of $1.324\pm0.177$ and $-3.679\pm0.161$  respectively derived by Cantat-Gaudin \etal (2018) for this cluster.

\begin{table*}
\tiny
\centering
\caption{ A sample table of the catalogue for the open cluster NGC 1513. Full table is available electronically. The (X,Y) pixel positions, RA and DEC sky coordinates and $U, B, V, R$ and $I$ CCD magnitudes with corresponding DAOPHOT errors are listed in columns 2 to 15 while GAIA EDR3 proper motion membership probability (MP) is given in the last column. }
\begin{tabular}{cccccccccccccccc}
\hline

    ID&X&Y&RA&DEC&U&eU&B&eB&V&eV&R&eR&I&eI&MP  \\
    &(pixel)&(pixel)&$J_{2000}$&$J_{2000}$&(mag)&(mag)&(mag)&(mag)&(mag)&(mag)&(mag)&(mag)&(mag)&(mag)&\\ \hline
    1&  531.257&   18.157& 4:10:07.96& 49:33:54.1&     16.031&  0.053& 15.394&  0.020& 14.600&  0.027& 14.113&  0.005& 13.604&  0.003&    72\\
    2&  597.253&   18.176& 4:10:06.66& 49:33:54.2&     99.999&  9.999& 99.999&  9.999& 21.765&  9.999& 20.308&  9.999& 19.189&  9.999&     0\\
    3&    6.203&   26.257& 4:10:18.27& 49:33:51.7&     20.859&  9.999& 22.205&  9.999& 99.999&  9.999& 99.999&  9.999& 99.999&  9.999&     0\\
    4&    4.954&   28.463& 4:10:18.30& 49:33:51.3&     99.999&  9.999& 20.844&  1.110& 18.971&  0.110& 99.999&  9.999& 99.999&  9.999&     0\\
    5& 1441.349&   30.289& 4:09:50.08& 49:33:53.0&     20.072&  0.093& 18.779&  0.013& 17.036&  0.003& 16.021&  0.007& 15.028&  0.017&     0\\
    -&-&-&-&-&-&-&-&-&-&-&-&-&-&-&-\\
    -&-&-&-&-&-&-&-&-&-&-&-&-&-&-&-\\
\hline
\end{tabular}
\end{table*}

\begin{figure}
\vspace{-1.4cm}
\begin{center}
\includegraphics[width=8cm, height=8cm]{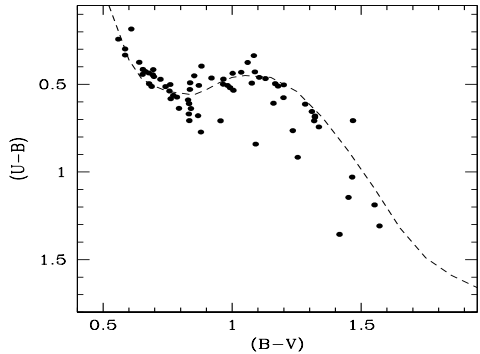}
\vspace{-2.1cm}
\caption{The $(U-B)$ vs $(B-V)$ colour-colour diagram using the cluster members. The dotted line represents
the locus of Schmidt-Kaler (1982) ZAMS overplotted on the observed stars. }
\label{red}
\end{center}
\vspace{-1.1cm}
\end{figure}

\subsection{Proper motion membership probability}

The cluster membership probability of a star is determined following the method described by Balaguer-N\'{u}\~{n}ez et al. (1998). For this, first proper motion frequency distributions of cluster stars ($\phi_c^{\nu}$) and field stars ($\phi_f^{\nu}$) are constructed by the equations given below:

\begin{center}
   $\phi_c^{\nu} =\frac{1}{2\pi\sqrt{{(\sigma_c^2 + \epsilon_{xi}^2 )} {(\sigma_c^2 + \epsilon_{yi}^2 )}}}$

$\times$ exp$\{{-\frac{1}{2}[\frac{(\mu_{xi} - \mu_{xc})^2}{\sigma_c^2 + \epsilon_{xi}^2 } + \frac{(\mu_{yi} - \mu_{yc})^2}{\sigma_c^2 + \epsilon_{yi}^2}] }\}$  and \\
$\phi_f^{\nu} =\frac{1}{2\pi\sqrt{(1-\gamma^2)}\sqrt{{(\sigma_{xf}^2 + \epsilon_{xi}^2 )} {(\sigma_{yf}^2 + \epsilon_{yi}^2 )}}}$

$\times$ exp$\{{-\frac{1}{2(1-\gamma^2)}[\frac{(\mu_{xi} - \mu_{xf})^2}{\sigma_{xf}^2 + \epsilon_{xi}^2}}
-\frac{2\gamma(\mu_{xi} - \mu_{xf})(\mu_{yi} - \mu_{yf})} {\sqrt{(\sigma_{xf}^2 + \epsilon_{xi}^2 ) (\sigma_{yf}^2 + \epsilon_{yi}^2 )}} + \frac{(\mu_{yi} - \mu_{yf})^2}{\sigma_{yf}^2 + \epsilon_{yi}^2}]\}$\\
\end{center}

where ($\mu_{xi}$, $\mu_{yi}$) are the proper motions of $i^{th}$ star, while ($\epsilon_{xi}$, $\epsilon_{yi}$) are the corresponding proper motion errors. ($\mu_{xc}$, $\mu_{yc}$) and  ($\mu_{xf}$, $\mu_{yf}$) represent the center of the cluster and field proper motion distributions respectively. For the cluster members, the intrinsic proper motion dispersion is denoted by $\sigma_c$, whereas $\sigma_{xf}$ and $\sigma_{yf}$ exhibit the field intrinsic proper motion dispersion. The correlation coefficient $\gamma$ is calculated as:

\begin{center}
$\gamma = \frac{(\mu_{xi} - \mu_{xf})(\mu_{yi} - \mu_{yf})}{\sigma_{xf}\sigma_{yf}}$.
\end{center}

To calculate $\phi_c^{\nu}$ and $\phi_f^{\nu}$, we considered those stars which have proper motion errors $\le$ 1 mas~yr$^{-1}$.  The center of the cluster proper motion is considered as mean proper motion derived in previous section.
The  intrinsic proper motion dispersion for the cluster stars ($\sigma_c$) could not be determined reliably using the present proper motion data. Assuming a distance of 1.33 kpc derived in the next section, and radial velocity dispersion 1 km/s for open clusters (Girard et al. 1989), the expected dispersion in proper motion would be $\sim$ 0.15 mas~yr$^{-1}$. For field stars, we have
($\mu_{xf}$, $\mu_{yf}$) = (0.5, -1.3) mas yr$^{-1}$ and
($\sigma_{xf}$, $\sigma_{yf}$) = (1.7, 1.6) mas yr$^{-1}$. The $n_{c}$ and $n_{f}$ are the normalized number of cluster and field stars respectively
(i.e., $n_c + n_f = 1$), the total distribution function can be calculated as:
\begin{center}
$\phi = (n_{c}~\times~\phi_c^{\nu}) + (n_f~\times~\phi_f^{\nu})$,  \\
\end{center}
Now, the membership probability (MP) for the $i^{th}$ star is derived as:
\begin{center}
$P_{\mu}(i) = \frac{\phi_{c}(i)}{\phi(i)}$. \\
\end{center}

The value of MP for each star is listed in Table 3 along with photometric and other relevant data. 

Earlier, Frolov et al. (2002) determined  proper motion of stars using the photographic plates. They considered the stars with proper motion error $\le$ 2 mas/yr for membership determination. There are 92 stars common between us and Frolov et al. (2002). We found 42 stars in Frolov et al. (2002) and 33 stars in our study which have a MP of more than 50\%. Because of the better accuracy of  proper motion data in GAIA EDR3 ($\le$ 1 mas/yr), the present estimate of MP is more reliable than the previous study. Out of 554 stars observed in the region of NGC 1513, only 106 stars have $MP \ge $ 50 \%. All these stars have been considered as cluster member and are used for estimating cluster parameter in the following sections.  

\begin{figure}
\vspace{-2.0cm}
\begin{center}
\includegraphics[width=9cm, height=10cm]{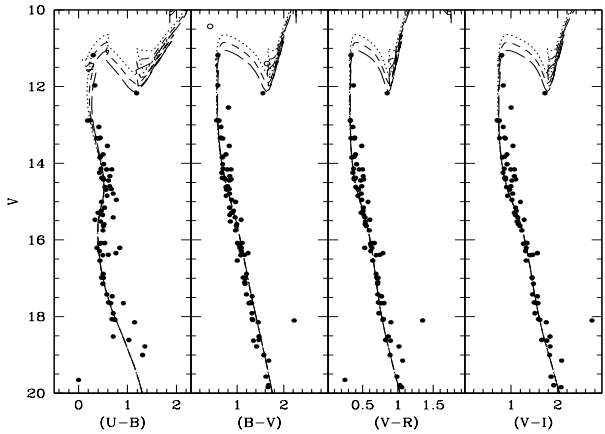}
\vspace{-3.0cm}
\caption{The $V$ vs $(U-B)$, $V$ vs $(B-V)$, $V$ vs $(V-R)$, and $V$ vs $(V-I)$ colour-magnitude diagrams of the cluster based on the stars having MP$>50\%$. The over plotted dotted, short dash and long dash lines are the isochrones of log(age) 8.30, 8.35 and 8.40 with $z=0.019$ taken from Girardi \etal (2000).}
\label{cmd1}
\end{center}
\vspace{-1.0cm}
\end{figure}
\section {The parameters of the cluster NGC 1513}

\subsection{Reddening}

To estimate the reddening towards the cluster region, we plot $(U-B)$ vs $(B-V)$ colour-colour diagram of the stars having membership probability more than 50\% in Fig. \ref{red}. The intrinsic ZAMS given by Schmidt-Kaler (1982) is fitted by considering the slope of reddening line as $E(U-B)/E(B-V) = 0.72$. Visual fitting of ZAMS to the colour-colour diagram provides a reddening value of  $E(B-V)=0.65\pm0.03$ mag for the NGC 1513. Present estimate of reddening is very close to the corresponding value 0.67 mag estimated by del Rio  \& Huestamendia (1988).

\subsection {Age and distance }

The photometric colour-magnitude diagrams (CMD) are used to estimate the age and distance of the cluster. The detail shape and different features in the CMDs are mainly depend on the age and metallicity of the cluster. The CMDs of GAIA EDR3 proper motion cluster members are shown in Fig. \ref{cmd1}. It is the deepest $V, (U-B); V, (B-V); V, (V-R)$ and $V, (V-I)$ CMDs extending up to $V \sim 21$ mag except in $V, (U-B)$ CMD where it is only up to $V \sim 19$ mag. A well defined cluster main-sequence is clearly visible in all the CMDs. Two bright stars, identified as proper motion cluster member by Frolov et al. (2002), but could not be observed by us, are also plotted in  Fig. \ref{cmd1}. One giant star has been saturated while the blue straggler is outside from our observed field of view. Their photometric data are borrowed from Frolov et al. (2002) and WEBDA catalogue of open star cluster. One of them is red giant in helium burning stage of stellar evolution while other one is possible blue straggler star. Another red giant, observed by us, is also a proper motion cluster member.

The age of the cluster is determined by fitting the theoretical stellar evolutionary isochrones of $Z=0.019$ taken from Girardi \etal (2000). The isochrones of log(age)=8.30, 8.35 and 8.40 are over plotted with dotted, short dash and long dash lines in all the CMDs. The overall visual fits of isochrones in the CMDs are satisfactory. The detailed shape of main-sequence and turn-off are reproduced. The best fitting isochrone to the main-sequence and turn-off provides an age of the cluster as 225$\pm$25 Myr (log(t)=8.35). The present age estimate is in good agreement with log(t)=8.4 derived by Frolov \etal (2002). The inferred apparent distance modulus $(m-M)=12.60\pm0.15$ mag provides a heliocentric distance as 1.33$\pm$0.10 kpc which is very close to the value of 1.32 Kpc derived by del Rio \& Huestamendia (1988). 
  
\section {The luminosity and mass function of NGC 1513}

For reliable estimation of luminosity and mass function of a star cluster, the first necessary step is to remove the field star contamination from the sample of stars. This has been done using proper motion MP derived in section 3.2. Only stars with MP $\ge$ 50\% are considered as cluster members. In order to avoid errors introduced due to data incompleteness, We have considered stars $V \le$ 19 mag only. The completeness is $\sim$ 100\% at the level of 19$^{th}$ mag (Fabricius et al. 2021).

\begin{figure}
\vspace{-1.0cm}
\begin{center}
\includegraphics[width=8cm, height=9cm]{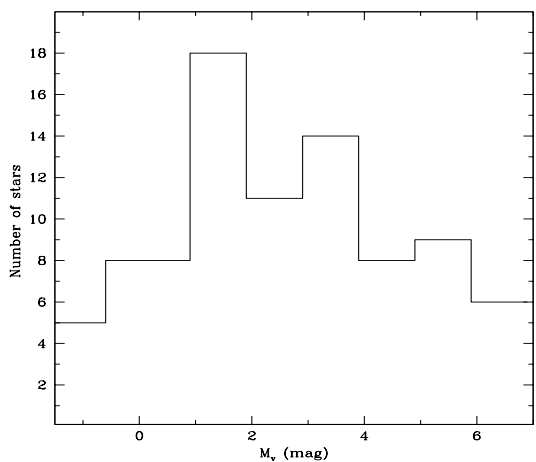}
\caption{The brightness distribution i.e. luminosity function of the open cluster NGC 1513.}
\label{lf}
\end{center}
\vspace{-1.0cm}
\end{figure}

\begin{figure}
\vspace{-1.5cm}
\begin{center}
\includegraphics[width=8cm, height=9cm]{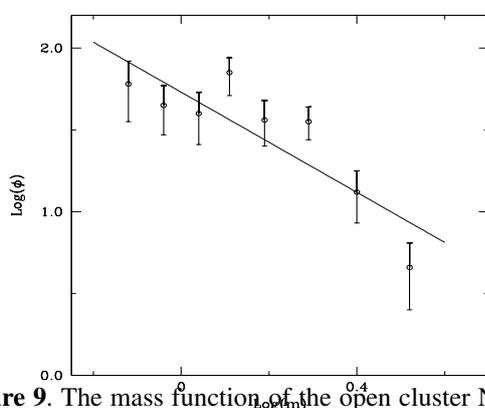}
\vspace{-3.0cm}
\caption{The mass function of the open cluster NGC 1513 derived using mass luminosity relation given by Girardi et al (2000) models.}
\label{mf}
\end{center}
\vspace{-1.0cm}
\end{figure}

To construct the luminosity function (LF), $V/(V-I)$ colour-magnitude has been used and the apparent $V$ magnitude is converted into the absolute magnitude ($M_v$) using the distance modulus of the cluster. The histogram of brightness distribution i.e. LF is shown in Fig \ref{lf}. The LF of the cluster rises until $M_v = 1.5$ mag and then decreases. Based on the cluster parameters derived in this study and theoretical models given by Girardi et al. (2000), we have converted LF in to mass function (MF) as shown in Fig \ref{mf}. The MF slope is estimated using the relation logd$N/dM = -(1+x)log(M) + $ constant. Here, d$N$ represents the number of stars in a mass bin d$M$ with central mass $M$ and $x$ is the slope of the MF. The MF slope derived in this analysis is $x=0.53\pm0.37$. Present estimate is very close to the value of  0.54$\pm$0.36, derived by Maciejewski et al. (2007). However, Our derived value is lower than the Salpeter (1955) value of 1.35 derived for field stars in the Solar neighbourhood.   

\section {The Galactic globular cluster NGC 4147}

The globular star cluster NGC 4147 ($\alpha_{2000}= 12^h 10^m 6.2^s; \delta_{2000}= +18^{\circ} 32^{\prime} 31^{\prime\prime}; l=253^{\circ},  b = +77^{\circ}$) is a halo object (Harris 2010). It is located near the boundary between the third and fourth Galactic quadrants, and not far from the north Galactic pole. The foreground reddening, $E(B-V)=0.02$ mag, is very small. This globular cluster is located at a distance of 17.3 kpc and 19.1 kpc from the Sun and the Galactic center respectively. This indicates that the cluster is a member of halo rather than disk population. The cluster metallicity ([Fe/H] = -1.83) listed in Harris (2010) catalogue is very low.

\begin{table*}
\centering
\caption{This is a sample table of the catalogue for the Galactic globular cluster NGC 4147. Full table is available electronically. The (X,Y) pixel positions, RA, DEC sky coordinates and the $B$ and $R$ CCD magnitudes with corresponding DAOPHOT errors are listed in columns 2 to 9 while the GAIA EDR3 proper motion MP is provided in last column. For most of the stars fainter than $V =$ 21 mag MP could not be estimated due to lack of proper motion data.}
\begin{tabular}{cccccccccc}
\hline

    ID&X&Y&RA&DEC&B&eB&R&eR&MP  \\
    &(pixel)&(pixel)&$J_{2000}$&$J_{2000}$&(mag)&(mag)&(mag)&(mag)&\\ \hline
    1&         670.600&    8.431&  12:10:02.27&   18:35:43.8&   20.072&    0.005&   18.256&    0.006&    0\\
2&         650.973&   15.857&  12:10:02.80&   18:35:40.9&   21.529&    0.018&   20.775&    0.021&    0\\
3&         640.832&   47.934&  12:10:03.07&   18:35:28.6&   20.573&    0.008&   19.873&    0.014&    0\\
4&         943.470&   51.257&  12:09:54.93&   18:35:28.2&   20.902&    0.012&   20.232&    0.016&    0\\
5&         763.679&   65.565&  12:09:59.76&   18:35:22.2&   21.158&    0.014&   20.421&    0.015&    0\\
    -&-&-&-&-&-&-&-&-&-\\
    -&-&-&-&-&-&-&-&-&-\\
\hline
\end{tabular}
\end{table*}

\subsection{Optical observations of NGC 4147}

The CCD images of the NGC 4147 were obtained with the 3.6-m DOT in standard Bessel $B$ and $R$ filters on 24$^{th}$ March 2017 using the CCD imager described above in section 2.1. Several images of 600 sec were taken in each $B$ and $R$ pass bands. Furthermore, observations were taken in 4$\times$4 pixel binning mode and all images in a filter were stacked together to improve the signal to noise ratio (S/N) for relatively fainter stars. The mean FWHM of the stars is 3.5 pixel (0.35 arcsec). The Bias and flat frames were acquired during the observations. These calibration frames were used to clean the science frames using the standard IRAF. The DAOPHOT/ALLSTAR (Stetson 1987, 1992) package was used to derive instrumental magnitudes. Quadratically varying PSFs were used to derive instrumental magnitudes. They were converted to the standard system using 375 Stetson standard stars present in the field of NGC 4147 \footnote{https://www.canfar.net/storage/vault/list/STETSON/Standards}. The $B$ and $R$ magnitudes of the stars obtained in this way are listed in Table 4 along with their pixel, RA and Dec coordinates. The mean error is $\sim$ 0.1 mag at 24$^{th}$ mag. The format of the data is shown in table 4 while entire data table consisting 1825 stars is available online. The data table can also be obtained from the corresponding author.

\subsection {The colour-magnitude diagram of NGC 4147}

\begin{figure}
\vspace{-1.5cm}
\centering
\hbox{\includegraphics[width=8.5cm]{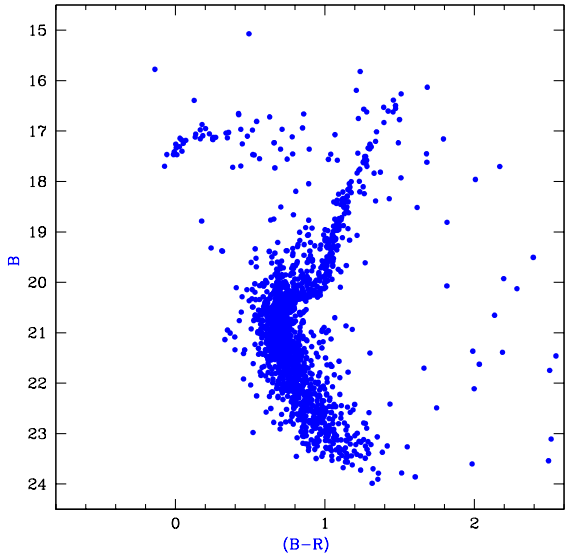}}
\vspace{-2.5cm}
\caption{The $B$, $(B-R)$ CMD of the globular cluster NGC 4147 using the data obtained with the 4K4$\times$4K CCD Imager mounted at the axail port of the 3.6-m DOT on 24-03-2017.}
\label{cmd2}
\vspace{-0.6cm}
\end{figure}

The $B$, $(B-R)$ CMD of NGC 4147 is shown in Fig. \ref{cmd2}. All the stars observed in $B$ and $R$ filter having an error better than 0.1 mag are shown in the figure. The main-sequence, giant branch and horizontal branch are clearly visible in the CMD. The main-sequence extends upto $\sim$ 24 mag towards the fainter end.

In order to estimate age and distance of the cluster, we identified GAIA EDR3 proper motion cluster members using procedure described above in Section 3.2. The ($\mu_{xc}$, $\mu_{yc}$) = (-1.7, -2.1) mas yr$^{-1}$ have been estimated from stars having proper motion error $\le$ 1 mas~yr$^{-1}$. The proper motion dispersion has been calculated as 0.03 mas~yr$^{-1}$ by considering a distance of 18.2 kpc derived in this section, and radial velocity dispersion 2.84 km/s taken from Baurgardt et al. (2018). For field stars, we have ($\mu_{xf}$, $\mu_{yf}$) = (-1.9, -2.2) mas yr$^{-1}$ and ($\sigma_{xf}$, $\sigma_{yf}$) = (1.6, 1.7) mas yr$^{-1}$.

\begin{figure}
 \vspace{-1.5cm}
\centering
\hbox{\includegraphics[width=8.5cm]{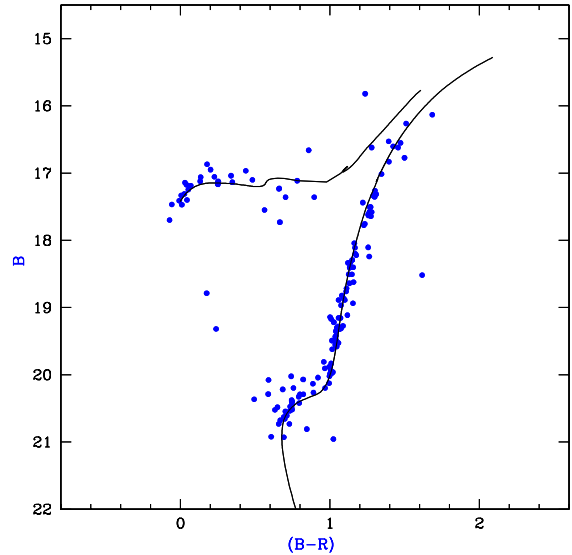}}
\vspace{-2.5cm}
\caption{The $B$, $(B-R)$ CMD of the globular cluster NGC 4147 using the stars having $Mp \ge$ 50\%. The solid line shown in the diagram is a 14 Gyr isochrone taken from BaSTI
(Hidalgo \etal 2018) for $Z=0.0003$.}
\label{cmdmp}
\vspace{-0.5cm}
\end{figure}

The age and distance of the cluster are estimated by visual fitting of the theoretical stellar evolutionary isochrone provided by Hidalgo \etal (2018) to the CMD. The $B$, $(B-R)$ CMD of 236  stars having $MP \ge$ 50\% is plotted in Fig. \ref{cmdmp}. We could, therefore, plot only stars with $V \le$ 21 mag. We superimposed the isochrone of age 14 Gyr with metallicity $Z=$ 0.0003 in Fig. \ref{cmdmp}. The fitting of isochrone is done with distance modulus $(m-M)=16.35$ mag and $E(B-R)=0.02$ mag. The estimated distance modulus provides a heliocentric distance of the cluster as 18.2 kpc which is identical to the value of 18.2 kpc derived by Baumgardt et al (2018). 

Optical identification of AstroSat discovered sources was one of the prime goals of the 3.6-m DOT and the 4K$\times$4K CCD Imager (Sagar 2018, Pandey et al.2018, Kumar \etal 2022). Recently, Kumar et al. (2021) identified several AstroSat-UVIT sources in the field of NGC 4147. These objects are located in different stages of stellar evolution in the UV-optical CMDs showing properties of blue horizontal branch stars, blue straggler star and variable stars. While Lata et al. (2019) have discovered several different types of variables within core region of the NGC 4147 using optical observations taken with the 4K$\times$4K CCD imager mounted on the 3.6-m DOT. However, only one object, located at RA = 182.527222 \& DEC = 18.55061 (J2000), is common between these two studies. This far-UV identification of an optical source within half light radius of the NGC 4147 shows properties of a RR Lyrae type of variables/eclipsing binary. The optical light curve of this common object is also shown in Fig 12. 

\begin{figure}
\vspace{-0.8cm}
\centering
\hbox{\includegraphics[width=8.5cm]{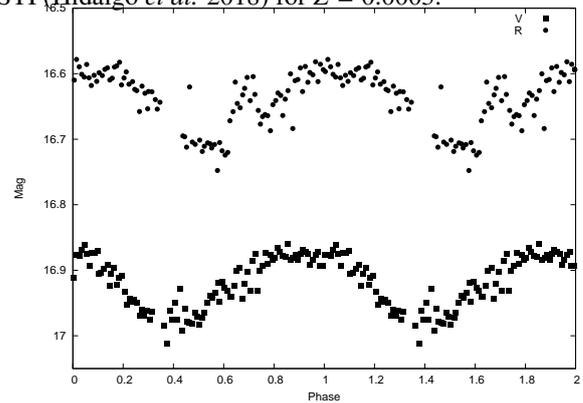}}
\vspace{-1.0cm}
\caption{$V$ and $R$ band light-curves of the star number V24 in Table 2 by Lata et al. (2019). It is numbered as HB24 in Table 4 of Kumar R. et al. (2021).}
\label{lc}
\vspace{-0.2cm}
\end{figure}

\section{Optical monitoring of white dwarf WD1145+017}

 The transiting behaviour of the dusty debris clouds orbiting around white dwarf WD 1145+017 was  discovered by Kepler {\it $K2$} mission (Vanderburg \& Rappaport 2018). Recent observations of WD 1145+017 are recently studied by Xu \etal 2018 (and references therein). These dust clouds reveal themselves through deep, broad, and evolving transits in the star's light curves. Such observations have shown transits with multiple periods ranging from 4.5 to 4.9 hours and transit duration ranging from $\sim$3 min to as long as an hour (Xu \etal 2018, Rappaport \etal 2018 and references therein). Here, we report CCD optical B pass band photometric observations of WD 1145+017  taken with the 3.6-m DOT on the night of 22-23 April 2017. Technical details of the CCD imager used by us are given above in section 2.1. The data was collected in the 4 $\times$ 4 binning mode with 1-MHz readout frequency (Kumar \etal 2022).  During 5.5 hours of our observations, a total of 210 continuous frames were acquired. Their exposure times ranged from 70 to 80 seconds. Along with target field, bias, dark and twilight flat field frames were also collected for calibration purpose. The basic image processing of the images was performed using IRAF. 
 
\begin{figure}
\centering
\hbox{\includegraphics[width=8.5cm]{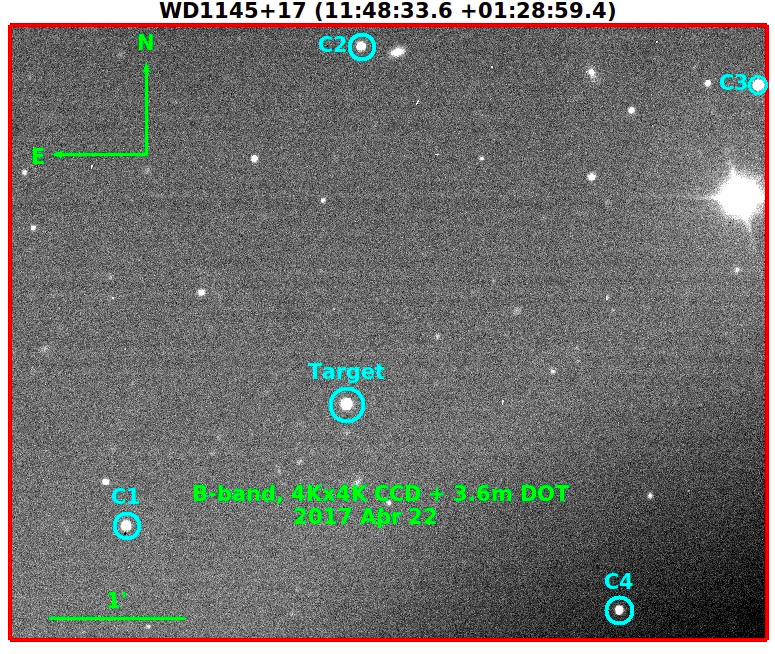}}
\caption{Finding chart of the WD1145+017 field showing the four comparison stars and the White Dwarf candidate as observed with the 3.6m DOT and 4K$\times$4K Imager in B-band.}
\label{FindingWhitedwarf}
\end{figure}

\begin{figure}[!t]
\includegraphics[scale=0.34]{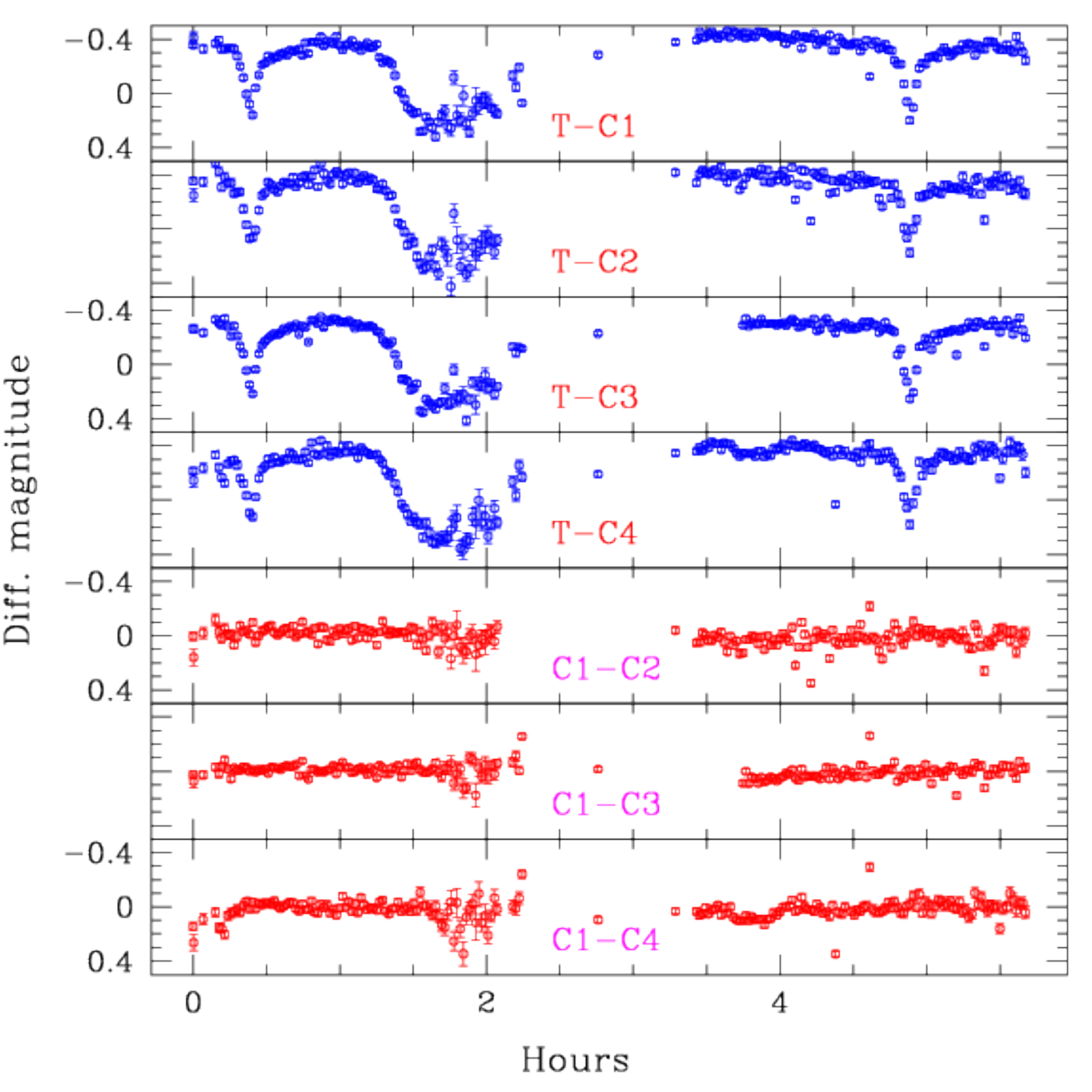}
\caption{Differential light curve of WD1145+017 with respect to all comparison stars are plotted in upper 4 panels (shown in blue colour) while differential magnitude of comparison stars C2, C3 and C4 with respect to star C1 are plotted against time of observations in lower 3 panels (red colour). Zero time of our $\sim$ 5.5 hours long observations corresponds to Julian date 2457866.109
i.e. 14:38:01 Universal time on 22 April 2017.}
\label{batlc1}
\end{figure}

Finding chart of the white dwarf is displayed in Figure \ref{FindingWhitedwarf}. In order to create differential light curves, we used 4 comparison stars identified as C1, C2, C3 and C4 in the  Figure \ref{FindingWhitedwarf}. All comparison stars are $\le  1$ magnitude fainter than the white dwarf. The FWHM of stellar images during our 5.5 hours of observations varied between $0\farcsec{8}$ to $1\farcsec{1}$. The instrumental magnitudes of the stars were obtained using the DAOPHOT package (Stetson 1987). In Figure \ref{batlc1}, differential magnitudes of W1145+017 with respect to all comparison stars are plotted against time of observations in upper 4 panels and shown with blue colour while differential magnitudes of comparison stars C2, C3 and C4 with respect to C1 are plotted against time of observations in lower 3 panels and shown with red colour. The DAOPHOT (or signal to noise) errors in our observations are $\sim$ 0.015 mag. There are no statistically significant variability observed in the comparison stars during entire period of our observations. While differential light curves of WD1145+017 exhibit three dips. They are similar with respect to all 4 comparison stars. Shape and depth of first and last dips are similar and separated by $\sim$ 4.6 hours. Since transit period of the objects revolving around the white dwarf ranges from 4.5 to 4.9 hours, both first and last dips of $\sim$ 20 minute duration correspond to the same orbiting body. It looks similar to dip A1 as reported by Xu \etal (2018) in Fig. discussing about their optical observations. Duration of the second dip is $\sim$ 80 minutes and light curve is quite complex in comparison to other two dips. Light curves with sharp decline and increase with shorter duration may indicate that orbiting object is solid and small in size while complex and longer duration light curve could be due to transit of dusty and larger sized body. A deeper analyses and modeling of these observations are in progress.  
     
\section{Deep high resolution images of Sh 2-61 in K band using the TIRCAM2 on 3.6-m DOT}

        \begin{figure*}
        \centering
        \includegraphics[width=0.49\textwidth, angle=0]{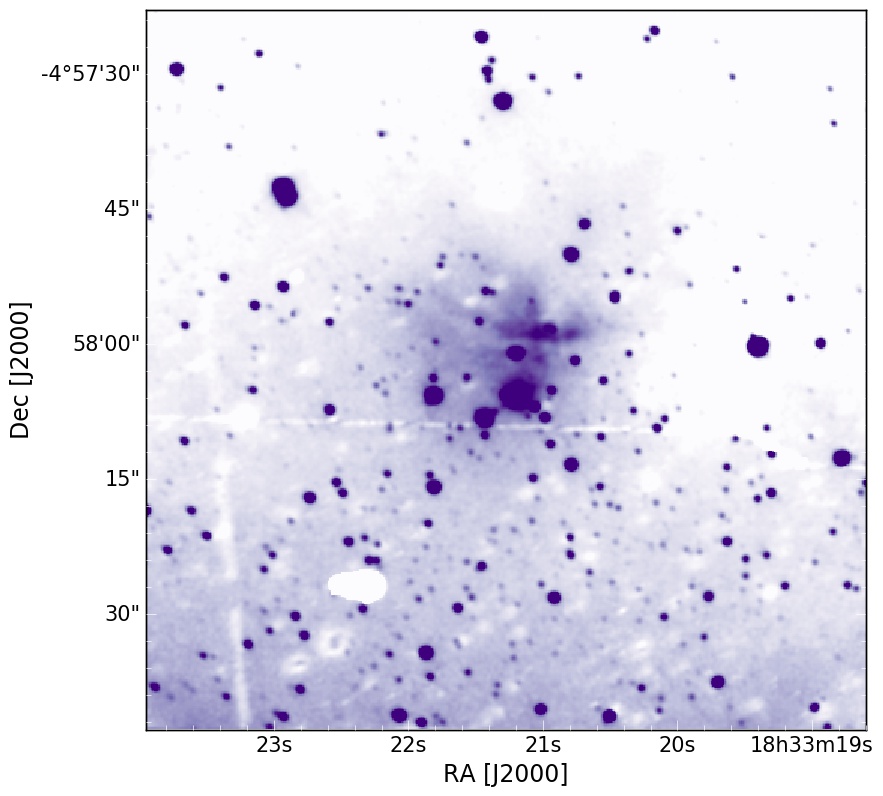}
        \includegraphics[width=0.49\textwidth, angle=0]{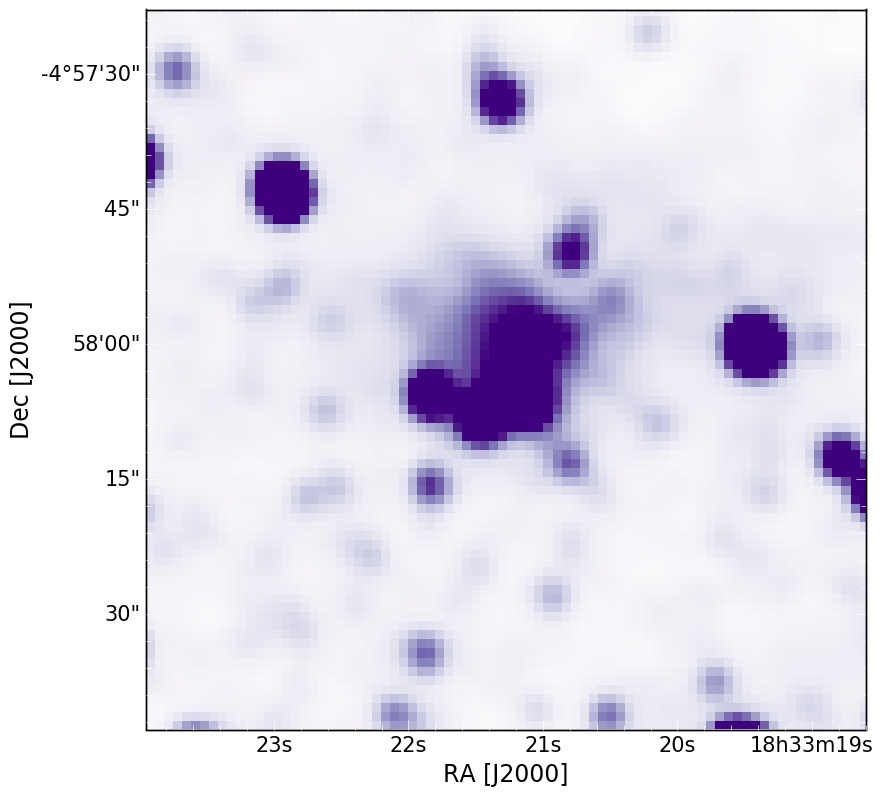}
        \caption{$K$-band image of Sh 2-61 region using the TIRCAM2 (left panel) and 2MASS (right panel) for a  $80\times80$ arcsec square FOV. }
        \label{img1}
        \end{figure*}

        \begin{figure*}
        \centering
        \includegraphics[width=0.27\textwidth, angle=0]{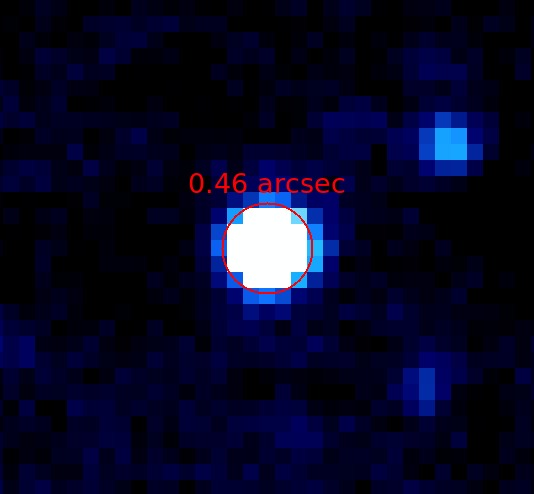}
        \includegraphics[width=5.3cm, height = 4.5cm, angle=0]{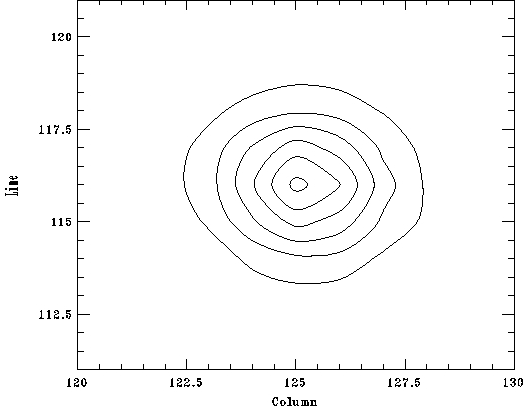}
        \includegraphics[width=0.36\textwidth, angle=0]{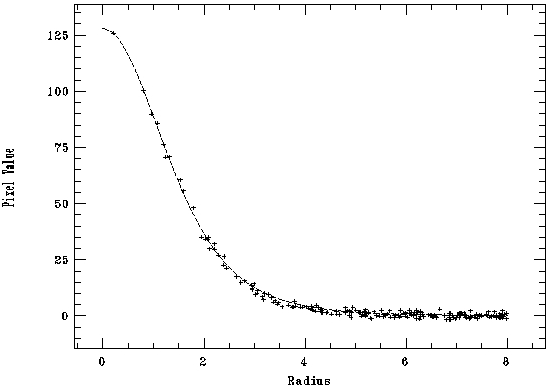}
        \caption{$K$-band image of a stellar profile (left panel),   intensity contours (middle panel), and radial profile (right panel) using the TIRCAM2 on 3.6m DOT. The image profile is a very sharp and is almost circular (ellipticity = 0.02). A circle of radius 0.46 arcsec (FWHM of PSF) is also shown in the left panel.}
        \label{img2}
        \end{figure*}

        \begin{figure*}
        \centering
        \includegraphics[width=0.46\textwidth, angle=0]{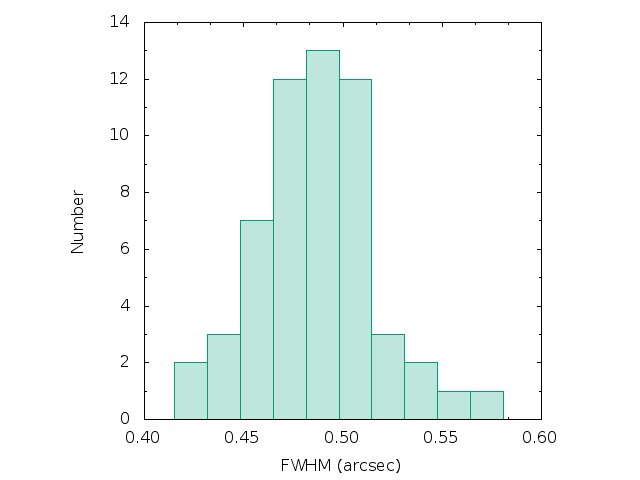}
        \includegraphics[width=0.49\textwidth, angle=0]{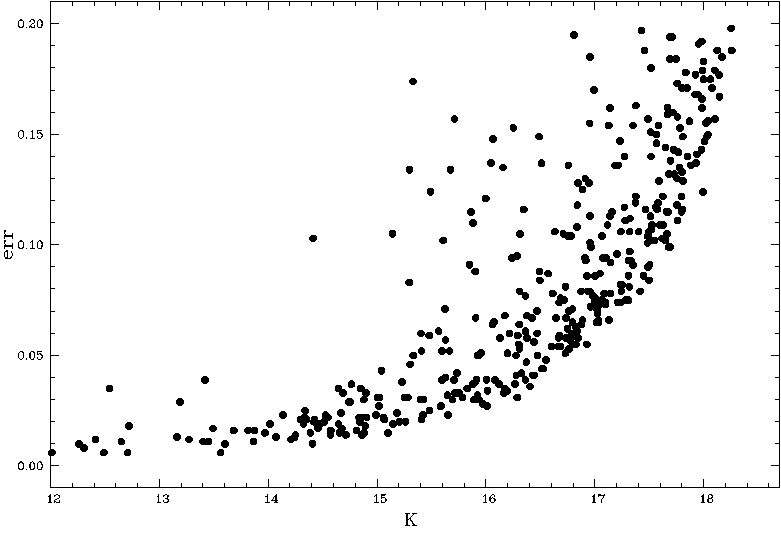}
        \caption{(Left panel): Distribution of the FWHM of the stellar sources in the TIRCAM2 field of view. The FWHM varies from 0.43 to 0.58 arcsec with peak around $\sim$0.5 arcsec. (Right panel): DAOPHOT error as a function of $K$-band magnitude. The source of $K\sim$18.2 mag is detected in $\sim$17 min exposure with error $\sim$0.2 mag (S/N $\sim$5).}
        \label{img3}
        \end{figure*}

Deep K-band (2.19 $\mu$m) photometric observations of the Sharpless region `Sh 2 -161' \footnote{http://galaxymap.org/cat/view/sharpless/61} $(\alpha_{J2000}: 18^h 33^m 21^s, \delta_{J2000} = -04^{\circ}58^{\prime} 02^{\prime\prime})$ were taken during the night of 2017 May 23 using TIRCAM2 (Naik \etal 2012; Baug \etal 2018) mounted at the Cassegrain main port of the 3.6-m DOT. The weather conditions in these nights were good with a relative humidity of $<$60 \% and the FWHM of the stellar images were $\sim$0\farcsec{5}. The field of view (FOV) of the TIRCAM2 is $\sim$86\farcsec{5} × 86\farcsec{5} square with a plate scale of 0\farcsec{167}. We took 101 frames in a dithered pattern around the central region of Sh2 -161 with exposure time of 10 s, i.e.  the total exposure time was $\sim$17 mins. Dark and sky flats were also taken during the observations. Sky frame was generated from the median of the dithered frames.  

The basic image processing such as dark/sky subtraction and flat-fielding was done using tasks available within IRAF. Final image was generated by aligning and combining the sky subtracted frames and is shown in the left panel of Fig. \ref{img1}. For  comparison, we have also shown the 2MASS image of the similar FOV. Clearly, using TIRCAM2 we can resolve far more stars than 2MASS and observe dust and gas emission near the central region of Sh 2-61. In Fig. \ref{img2}, we have demonstrated the sharp image quality of the TIRCAM2 data as evident from the low ellipticity ($\sim$0.02) and FHWM ($\sim$0\farcsec{46}) of a stellar image.  In the left panel of Fig. \ref{img3}, we have shown the distribution of FWHM of the stellar profile in the final image. The distribution also points towards a good quality image from TIRCAM2 as the FWHM varies from 0\farcsec{43} to 0\farcsec{58} with median $\sim$ 0\farcsec{5}. The Instrumental magnitudes from the final image were obtained using the DAOPHOT package. As the region is crowded, we carried out PSF photometry to get the magnitudes of the stars. Astrometry of the stars was done using the Graphical Astronomy and Image Analysis Tool\footnote{http://star-www.dur.ac.uk/~pdraper/gaia/gaia.html} with an rms noise of the order of $\sim$0\farcsec{28}. The calibration of the photometry to the standard system was done using the offset calculated from the 2MASS data. The typical DAOPHOT errors as a function of corresponding standard magnitudes are shown in the right panel of Fig. \ref{img3}. In total, 405 stars were identified in the central region of the Sh 2-61 with error $<$ 0.2 mag with detection upto K $\sim$ 18.2 mag. 2MASS has detected only 67 stars in the same FOV. All these clearly demonstrate the deep and high resolution imaging capability of TIRCAM2 on 3.6-m DOT in NIR bands.

\section{Summary and Conclusions}

Optical observations of an open cluster NGC 1513, a galactic globular cluster NGC 4147 and a white dwarf WD1145+07 and K-band imaging of a star forming region Sh 2-61 with 3.6m DOT are presented here. These observations were taken between November 2016 to May 2017. The main findings of the present analysis are as follows.

\begin{itemize}

\item The GAIA EDR3 proper motion data are used to identify 106 likely members in the open star cluster NGC 1513 having mean proper as $\mu_{\alpha}Cos{\delta}=1.29\pm0.02$ and $\mu_{\delta}=-3.74\pm0.02$ mas yr$^{-1}$.\\

\item Using the two colour diagram, we estimated reddening $E(B-V)$ as 0.65$\pm$0.03 mag to the NGC 1513. Distance to the cluster is determined as $1.33\pm0.10$ kpc while age is estimated as $225\pm25$ Myr by comparing the cluster's sequence with the theoretical isochrones of solar metallicity given by Girardi et al. (2000).\\

\item Based on the $B$, $(B-R)$ CMD of the GAIA EDR3 proper motion members, We determined the distance and age of the galactic globular cluster NGC 4147 as 18.2 kpc and 14 Gyr respectively.

\item The values of FWHM estimated from images of stellar sources are expected to be slightly more than the value of atmospheric seeing prevailing at the epoch of observations (Sagar \etal 2020). Atmospheric seeing is $\lambda$ dependent and varies as $\lambda^{-0.2}$. Observations reported here and earlier by Sagar \etal (2019, 2020) and recently by Panwar \etal (2022) show that at Devasthal, for a good fraction of observing time, optical to NIR sky images with sub-arc-sec resolution can be obtained with the 3.6-m DOT. These good sky conditions enable deep imaging of stars with detection of $B = 24.5\pm0.2, R = 24.6\pm0.12$ and $ V =25.2\pm0.2$ mag stars in exposure time of 1200, 4320 and 3600 second respectively. The NIR observations taken with TIRCAM2 show that stars up to $J = 20\pm0.1,  H = 18.8\pm0.1$ and $K = 18.2\pm0.1$ mag can be detected in effective exposure times of 500, 550 and 1000 second respectively. 

 \item The modern active optics 3.6-m DOT observing facilities have started providing good quality of optical and NIR observations for a number of front line Galactic and extra-galactic astrophysical research problems including optical follow up of $\gamma$-ray, X-ray, UV and radio sources observed with facilities like GMRT and AstroSat etc. and optical transients objects like SN and afterglows of $\gamma$-ray bursts and gravitational wave etc. The telescope has therefore started contributing significantly to the growth of our understanding of Astrophysical jets.  

\end{itemize}
\section*{Acknowledgements}
This manuscript is based on an invited talk delivered during international workshop Astrophysical jets and observational facilities: National perspective held by ARIES during April 5 to 9, 2021. One of us (Ram Sagar) thanks the National Academy of Sciences, India (NASI), Prayagraj, for the award of a NASI Honorary Scientist position; the Alexander von Humboldt Foundation, Germany, for the award of Group linkage long-term research program; and the Director, IIA, for hosting and providing infrastructural support during this work. SBP acknowledges financial support received from the BRICS grant {DST/IMRCD/BRICS/PilotCall1/ProFCheap/2017(G)}. SJ thanks to Nand Kumar and Amit Kumar for their helps in data analysis and preparing the figures. The 3.6-m Devasthal Optical Telescope (DOT) is a National Facility run and managed by Aryabhatta Research Institute of Observational Sciences (ARIES), an autonomous Institute under Department of Science and Technology, Government of India. This research has made use of data obtained from the High Energy Astrophysics Science Archive Research Center (HEASARC) and the Leicester Database and Archive Service (LEDAS), provided by NASA's Goddard Space Flight Center and the Department of Physics and Astronomy, Leicester University, UK, respectively. This work has made use of data from the European Space Agency (ESA) mission GAIA processed by Gaia Data processing  and Analysis Consortium
(DPAC), (https://www.cosmos.esa.int/web/gaia/dpac/consortium). It is worthy to mention that, this work has also used WEBDA.



\end{document}